\documentclass[12pt]{article}
\usepackage{amssymb,theorem,amsmath}
\usepackage[dvips]{graphicx}
\let\Prp=\Pr \def\Pr{\Prp\nolimits}
\newcommand{\be}{\begin{equation}}
\newcommand{\ee}{\end{equation}}

{\theorembodyfont{\rm}

 }

\newcommand{\cfig}[1]{Figure~\ref{#1}}

    \def\bbz{\mathbb{Z}}

\def\Prob{\mathop{\rm Prob}}

\def\n{{\bf n}}

\def\muhat{\hat\mu}\def\rhohat{\hat\rho}
\def\murho{\mu^{(\rho)}}

\def\interiorM{\rlap{\raise9pt\hbox to9.5pt{\hss$\scriptscriptstyle\circ$}}M}
\long\def\kill#1\endkill{\relax}

\def\0{{\it0}}\def\1{{\it1}}\def\2{{\it2}}\def\3{{\it3}}

\newdimen\mbsize \mbsize=\hsize \multiply\mbsize by 17  \divide\mbsize by 20

\numberwithin{equation}{section}
\numberwithin{theorem}{section}

\begin{document}

\title{\vskip-0.2truein
 Exact Solution of the F-TASEP}
\author{S. Goldstein\footnote{Department of Mathematics,
Rutgers University, New Brunswick, NJ 08903.},
J. L. Lebowitz\footnotemark[1],
\footnote{Also Department of Physics, Rutgers.}\ \ 
and E. R. Speer\footnotemark[1]}
\date{April 16, 2019}
\maketitle


\begin{abstract} We obtain the exact solution of the facilitated totally
asymmetric simple exclusion process (F-TASEP) in 1D.  The model is
closely related to the conserved lattice gas (CLG) model and to some
cellular automaton traffic models.  In the F-TASEP a particle at site $j$
in $\bbz$ jumps, at integer times, to site $j+1$, provided site $j-1$ is
occupied and site $j+1$ is empty.  When started with a Bernoulli product
measure at density $\rho$ the system approaches a stationary state. This
non-equilibrium steady state (NESS) has phase transitions at $\rho=1/2$
and $\rho=2/3$. The different density regimes $0<\rho<1/2$,
$1/2<\rho<2/3$, and $2/3<\rho<1$ exhibit many surprising properties; for
example, the pair correlation $g(j)=\langle\eta(i)\eta(i+j)\rangle$
satisfies, for all $n\in\bbz$, $\sum_{j=kn+1}^{k(n+1)}g(j)=k\rho^2$, with
$k=2$ when $0\le\rho\le1/2$, $k=6$ when $1/2\le\rho\le2/3$, and $k=3$
when $2/3\le\rho\le1$. The quantity $\lim_{L\to\infty}V_L/L$, where $V_L$
is the variance in the number of particles in an interval of length L,
jumps discontinuosly from $\rho(1-\rho)$ to 0 when $\rho\to1/2$ and when
$\rho\to2/3$.  \end{abstract}

\section{Introduction\label{intro}}

Phase transitions with absorbing states are a well-established field of
study \cite{MD}.  In particular, the facilitated time evolution of
particles on a lattice has been the subject of extensive study in both
the physics and mathematics literature.  In two or more dimensions, much
is known from numerical simulation \cite{rpv,hl,mcl}, but there are few
analytic results (but see \cite{ST}); in one dimension both numerical and
analytic studies are available \cite{oliveira,gkr,bbcs}.  Here we present
exact results for the {\it facilitated totally asymmetric simple
exclusion process} (F-TASEP); for more details and proofs, see
\cite{gls}. The F-TASEP is a variation of the conserved lattice gas model
of particles moving on the one dimensional lattice $\bbz$.  If we think
of the empty sites as cars and the occupied sites as empty spaces then
the model corresponds to one of the traffic models considered in
\cite{GG,LZGM}, as we discuss later in this section.

A configuration of the model is an arrangement of particles on $\bbz$,
with each site either empty or occupied by a single particle, that is,
the configuration space is $\{0,1\}^\bbz$; we write
$\eta=(\eta(i))_{i\in\bbz}$ for a given configuration.  We study the
discrete time dynamics, in which every particle whose left-hand
neighboring site is occupied attempts, at each integer time, to jump to
the neighboring site on its right; the jump takes place if the target is
unoccupied.  The model is thus a deterministic cellular automaton.  We
let $\eta_t$ denote the configuration at time $t$; the evolution is
deterministic and we can determine the ultimate fate of any initial
configuration with a well defined density $\rho$ \cite{gls}.  We will
describe the translation invariant (TI) states of the system, at any
density $\rho\in[0,1]$, which are stationary under the dynamics (the TIS
states).  In particular we will determine, completely or (in one phase)
partially, the final TIS state when the system is started in a
Bernoulli measure: an initial state $\murho$ in which each site is
independently occupied with probability $\rho$.

We will make use of a closely related model, the {\it totally asymmetric
stack model} (TASM), another particle system on $\bbz$ evolving in
discrete time.  In the TASM there are no restrictions on the number of
particles at any site, so that the configuration space is
$\{0,1,2,\ldots\}^\bbz$.  We denote stack configurations by boldface
letters.  The dynamics is as follows: at each integer time, every stack
with at least two particles ($\n(i)\ge2$) sends one particle to the
neighboring site to its right.  There is a natural but somewhat loose
correspondence of this model with the F-TASEP, with a stack configuration
$\n$ corresponding to a particle configuration in which strings of
$\n(i)$ particles are separated by single holes (a similar correspondence
is often introduced for zero range processes, see e.g. \cite{eh}).  At
the configuration level this correspondence is not one to one, but one
may show that it gives rise to a bijective correspondence between the TI
states, and also between the TIS states, of the two models \cite{gls2}.
In particular, if $\hat\mu$ is a TI state for the TASM, with density
$\hat\rho$, then the corresponding state $\mu$ of the F-TASEP has density
$\rho=\hat\rho/(1+\hat\rho)$.  If $\mu=\murho$ then in the corresponding
TASM measure $\muhat=\muhat^{(\rho)}$ the $\n(i)$ are i.i.d.~with
geometric distribution: $\muhat^{(\rho)}\{\n(i)=k\}=(1-\rho)\rho^k$.

We note some simple properties of the dynamics in the TASM.
Letting $\n_t(k)$ denote the height at time $t$ of the stack of particles
on site $k$, we have 

\begin{itemize}
\item If $\n_t(k)\ge2$ then $\n_{t+1}(k)=\n_t(k)$ unless $\n_t(k-1)\le1$,
  in which case $\n_{t+1}(k)=\n_t(k)-1;$

\item If $\n_t(k)\le1$ then $\n_{t+1}(k)=\n_t(k)$ unless $\n_t(k-1)\ge2$,
  in which case $\n_{t+1}(k)=\n_t(k)+1.$
\end{itemize}

Thus the possible changes in the value of $\n(k)$ in one step of the
dynamics, say from $t$ to $t+1$, may be summarized as
 \be\label{shel}
0\,\rightarrow\,1\,\leftrightarrows\,2\,\leftarrow\,3\,\leftarrow
   \,4\,\leftarrow\,5\,\leftarrow\,\cdots.
 \ee
 The indicated increases occur if and only if $\n_t(k-1)\ge2$, and the
decreases if and only if $\n_t(k-1)\le1$.

 Suppose now that $\muhat$ is a TIS state for the TASM, with density
$\rhohat$.  The discussion above then implies that at most three types of
stack configurations can occur with nonzero probability in $\muhat$:
(a)~all stacks have height 0 or 1; (b)~all stacks have height 1 or 2;
(c)~all stacks have height 2 or more.  This follows from the observation
that any stack of size 0 or 1 will increase if it is preceded by a stack
of size 2 or more and any stack of size 3 or more will decrease if it is
preceded by a stack of size 0 or 1, while according to \eqref{shel} no
new stacks of size 0 or of size 3 or more can be formed.

Now, given $\muhat$ as abov, let $\mu$ denote the corresponding TIS state
of the F-TASEP and let $\rho$ be the particle density in $\mu$.  The
possibilities discussed above then become: (a)~no two adjacent sites are
occupied by particles; (b)~no two consecutive sites are empty and no
three consecutive sites are occupied; (c)~no two sites which are adjacent
or at a distance of 2 from each other are both empty.  It is now obvious
that case (a) occurs when $\rho\le1/2$ ($\rhohat\le1$), (b) when
$1/2\le\rho\le2/3$ ($1\le\rhohat\le2$), and (c) when $\rho\ge2/3$
($\rhohat\ge2$).  At density $\rho=1/2$ there is a unique TIS state of
the F-TASEP, the superposition of the two (stationary but not TI) states
concentrated on the two periodic configurations with alternating 1's and
0's, while in the TASM there is a unique TIS state, with density
$\rhohat=1$ and $\n(j)=1$ for all $j$.  Similarly at $\rho=2/3$ the
F-TASEP has a unique TIS state, the superposition of three states which
are periodic with repeating unit $1\,1\,0$, and the TASM at $\rhohat=2$
has a unique TIS state, with $\n(j)=2$ for all $j$.

We will refer to the regions $0<\rho<1/2$, $1/2<\rho<2/3$, and
$2/3<\rho<1$ as the low, intermediate, and high density regions,
respectively.  In these regions the dynamics of the F-TASEP takes, with
probability one in any stationary state, a simple form: in the low
density region, configurations do not change with time; at intermediate
density, configurations translate two sites to the right at each time
step; and at high density, configurations translate one site to the left
at each time step. The corresponding result for the stack model is that
configurations in the low or high density region are fixed under the
dynamics, and those in the intermediate region translate one site to the
right at each time step.

By considering sites occupied by particles as empty, and empty sites as
occupied by cars, we can think of the F-TASEP as a traffic model in which
cars advance synchronously whenever there are at least two empty spaces
ahead of them.  This is a special case of the family of traffic models
considered in \cite{GG}, obtained by taking $\alpha=\delta=1$ and
$\beta=\gamma=0$ in the classification of Table~1 of that reference.  The
three phases found above correspond then to three patterns of traffic
flow; the phase boundaries for our case were determined, and the phases
described, in \cite{GG}.  Our high density region, $\rho>2/3$, now
becomes a low density car region, $\tilde \rho=1-\rho<1/3$, in which
traffic flows freely, with each car moving at the maximal possible
velocity one.  Our low density region, $\rho<1/2$, becomes a jammed
region, with $\tilde\rho>1/2$, in which there is no motion.  Finally, in
our intermediate density region stretches of jammed traffic, in which
each car sees only one empty site in front of it and therefore cannot
move, alternate with stretches in which each car sees two empty spaces
and hence moves with velocity one.

We now turn to a more detailed discussion of the behavior in each region
of density.  As indicated above, we are particularly interested in
studying the distribution in the limit $t\to\infty$ when the initial
configuration has a Bernoulli distribution.  The existence and uniqueness
of such a final distribution, for $\rho\notin\{1/2,2/3\}$, is easy to
show; the analysis for $\rho=1/2$ and $\rho=2/3$ is more complicated.
See \cite{gls}.

\section{The low density region}\label{low}

We now consider the case in which the initial configuration $\eta_0$ is
distributed according to the product Bernoulli measure $\murho$, with
$0<\rho<1/2$, and want to find the final (stationary) distribution of
$\eta$.  We will denote expectation values in the final measure by
$\langle\cdot\rangle$.  We have already observed that this measure will
be supported on trapped configurations with no adjacent occupied sites,
and what we want to know is the probability distribution of these
configurations.  An important observation \cite{gls} is that almost every
initial configuration $\eta_0$ in fact evolves to a final configuration
$\eta_\infty$.

 A little thought shows that any two consecutive sites which at some time
during the evolution are not both empty will then be nonempty at all
later times, so that two consecutive empty sites in any final
configuration $\eta_\infty$ must have been empty at all times.  This
means that no particle has departed from these sites during the evolution
and so the distribution of $\eta_\infty$ to the right of a double zero is
independent of what preceded, i.e., what follows is the same as if we
started with a semi-infinite Bernoulli measure at density $\rho$.  The
stationary state is thus given by a renewal process \cite{Feller2}, and
all we need to know for its complete description is the probability that
a double zero is followed by exactly $n$ pairs $1\,0$ (and then an
additional 0).  This probability is given by $c_n\rho^n(1-\rho)^{n+1}$,
where
 \be\label{catalan2}
c_n=\frac1{n+1}\binom{2n}n,\quad n=0,1,2,\ldots;
 \ee
 is the $n^{\rm th}$ {\it Catalan number} \cite{Catalan}; $c_n$ counts
the number of strings of $n$ 0's and $n$ 1's in which the number of 0's
in any initial segment does not exceed the number of 1's.  We remark that
the density of double zeros is given by
$\langle(1-\eta(j))(1-\eta(j+1))\rangle=1-2\rho$, since
$\langle\eta(j)\eta(j+1)\rangle=0$; this will be used in \eqref{surprise}
below.

The structure of the TIS state for $\rho\le1/2$, that is,
$\rhohat\le1$, is particularly simple for the TASM: $\n(i)$ can take only
values 0 and 1, and a double zero in the F-TASEP corresponds to a zero in
the TASM.  We may thus write $\zeta(i)=1-\n(i)$ and view
$\zeta\in\{0,1\}^\bbz$ as a lattice gas configuration.  Then given that
$\zeta(j)=1$, the probability that the {\it next} particle to the right
of $j$ will be at $j+k$ ($k>0$) is independent of $j$ and of $\zeta(j')$,
$j'<j$, and is given by $c_{k-1}\rho^{k-1}(1-\rho)^k$.

By calculating the generating function of the pair correlation function
$g(k)=\langle\eta(j)\eta(j+k)\rangle$ in the low density TIS state
of the F-TASEP, which may be obtained from the generating function of the
distance between two successive double zeros, i.e., of the Catalan
numbers, we discover an unexpected property of this state: for all
$n\ge0$.
 \be\label{lowid}
g(2n+1)+g(2n+2)=2\rho^2.
 \ee
  Alternatively, this can be obtained from another surprising property of
the stationary state: in the distribution $\langle\cdot\rangle_{00}$
defined by conditioning on the event that $\eta_{-1}=\eta(0)=0$, i.e., in
the stationary state for the semi-infinite system mentioned above,
$\langle\eta(2n+1)\rangle_{00}=\rho$ for any $n\ge0$, that is, the
probability for site $2n+1$ to be occupied is always $\rho$.  This is
clearly the case for $n=0$, and a simple but rather long inductive
analysis verifies it in general.  But then for any $n\ge0$,
 \begin{align}\nonumber
  \rho&= \langle\eta(2n+1)\rangle_{00}\\
  &=\frac1{1-2\rho}\langle(1-\eta(-1))
    (1-\eta(0))\eta(2n+1)\rangle\nonumber\\
  &=\frac1{1-2\rho}(\rho-g(2n+1)-g(2n+2))\label{surprise},
 \end{align}
 and this implies \eqref{lowid}

We can also show that the generating function of the truncated two point
function $g^T(k):=g(k)-\rho^2$ has radius of convergence
$\alpha=(4\rho(1-\rho))^{-1/2}>1$, so that $g^T(k)$ decays as
$\alpha^{-k}$, up to possible prefactors $k^\beta$.  From \eqref{lowid}
it now follows that, in the stationary state, the asymptotic value of
$V_L/L$, where $V_L$ is the variance of the number of particles in an
interval of length $L$, has the same value as for the initial Bernoulli
measure: by a simple argument,
 \be
\frac{V_L}L=\rho-\rho^2+\sum_{k=1}^\infty g^T(k)+\frac{o(L)}L,
 \ee
 and \eqref{lowid} implies that 
 \be
\sum_{k=1}^n g^T(k)=\left.\begin{cases}g^T(n),&\text{if $n$ is odd,}\\
  0,&\text{if $n$ is even,}\end{cases}\right\}\quad\longrightarrow\quad0.
 \ee
 Hence $v:=\lim_{L\to\infty}V_L/L=\rho(1-\rho)$ for $\rho<1/2$.

\begin{figure}[ht!]  \begin{center}
\includegraphics[width=5.0truein,height=2.0truein]{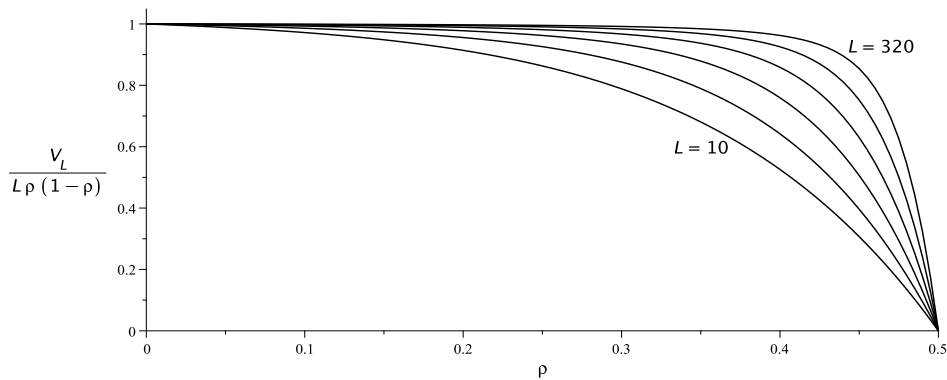} 
\caption{The normalized variance per unit length versus $\rho$, for
  $L=10$, 20, 40, 80, 160, and 320.}
\label{fig1} 
\end{center}
\end{figure}

At $\rho=1/2$, on the other hand, $V_L$ is of order 1 in the TIS state,
since the system is ordered (as $\ldots1\,0\,1\,0\,1\,0\ldots$) in one of
two possible configurations.  We thus have a sharp transition in $v$.
For a finite but large $L$ there is a transition region, determined
numerically to have width $L^{-1/2}$; see \cfig{fig1}.  A very similar
phenomenon occurs when $\rho=2/3$.

\bigskip

There is a different way to see that $v=\rho(1-\rho)$ and that in fact
this relation holds for all times, not just in the final state.  For each
particle lies initially in an interval between two double zeros which
will persist into the stationary state, and the distance the particle can
travel from $t=0$ to $t=\infty$ is bounded by the length of this
interval, which has expected value $1/(1-2\rho)$. Thus the number of
particles in an interval of large length $L$ will, at any time $t$,
differ little from the number at $t=0$.  We can in fact compute exactly
the distribution of the distance $d$ that a typical particle travels from
its initial to its final position:
 \be
\Prob(d=k)= \frac{1-2\rho}{1-\rho}\left(\frac\rho{1-\rho}\right)^k.
 \ee
 The average distance moved is $\rho/(1-2\rho)$.  Note that as
$\rho\nearrow1/2$ this expected distance diverges.  This is so because at
$\rho=1/2$ the infinite system never becomes fixed, but rather oscillates
(locally) between the two stationary configurations.

\section{The high density region, $2/3<\rho<1$}\label{high}

As shown earlier, the only configurations possible in the stationary
state at high densities are those in which there are no adjacent zeros
or zeros separated by one occupied site.  In terms of the stack model
this corresponds to all sites being occupied by two or more particles:
$\n(k)\ge2$ for all $k$.

To bring out the parallels between the behaviors of the model in the low
and high density regions it is convenient here to work in a moving frame.
Thus in this section we consider a modified F-TASEP dynamics for which at
each time step one first executes the F-TASEP rule, then adds a
translation by one lattice site to the right, so that the possible
limiting configurations described above are stationary under the new
dynamics.  No change is necessary for the TASM dynamics.  Under the
modified F-TASEP dynamics almost every initial configuration $\eta_0$ for
the measure $\murho$ will in fact evolve to a final configuration
$\eta_\infty$, and the same conclusion holds for the TASM.

The TIS state obtained from a high-density initial Bernoulli distribution
has properties similar to those of the low density TIS state.  In
particular, the role of double zeros in defining a renewal
process in the low density state is now played by blocks of three
consecutive occupied sites.  Since such a block may be followed by
additional occupied sites, it corresponds in the TASM to a site $k$ with
$\n(k)\ge3$ in the final configuration, and it follows from \eqref{shel}
that such a site must have contained three or more particles at all
times.  Hence it must have sent one particle to its right at each time
step, $t=0,1,2,\ldots$, so what happens to its right is independent of
what is to its left (given that what happens to its left is such that it
is such a site).  This implies a similar statement in the F-TASEP with
modified dynamics, and thus the final state is again obtained from a
renewal process, as at low density, albeit a more complicated one:
configurations in the final state have the form
 \be\label{highetainf}
\eta_\infty=\ldots\,(0\,1\,1)^{n_1}\,1\,(0\,1\,1)^{n_2}\,1\,\ldots,
 \ee
 where the $n_k$ are independent with a computable distribution.

With this in hand we can obtain various properties of the stationary
state which are analogous to those obtained in the low density region.
For example: (a)~if we condition on sites $-2$, $-1$, and $0$ being
occupied then $\langle\eta(3n+1)\rangle=\rho$ for any $n\ge0$,
(b)~$g(3n+1)+g(3n+2)+g(3n+3)=3\rho^2$ for any $n\ge1$ (compare
\eqref{lowid}), and (c)~$v=\rho(1-\rho)$.

\section{The intermediate density region}\label{inter}

The TIS states in the intermediate region have a more complex structure
than those in the high and low density regions.  The configurations
permitted in such states are those which do not contain any double zeros
or any blocks of consecutively occupied sites of length greater than two;
equivalently, these configurations are constructed from an ``alphabet''
consisting of the strings $01$ and $011$, that is, have the form
 \be\label{midetainf}
 \cdots(0\,1)^{n_k}(0\,1\,1)^{m_k}(0\,1)^{n_{k+1}}
    (0\,1\,1)^{m_{k+1}}\cdots,
 \ee
 In the TASM language these become configurations in which all sites are
occupied by either one or two particles.  To make configurations in the
stationary measure actually be stationary we shift each configuration,
after carrying out the prescribed forward jumps, to the left: by two
sites in the F-TASEP and one site in the TASM.  Under the modified
F-TASEP dynamics almost every initial configuration $\eta_0$ for the
measure $\murho$ will in fact evolve to a final configuration
$\eta_\infty$, and the same conclusion holds for the TASM.

When the F-TASEP is started in an initial Bernoulli measure the final
state is no longer a renewal process, at least so far as we can tell.  In
this case we cannot compute either the distribution of the variables
$n_k$ and $m_k$ of \eqref{midetainf} or the pair correlation function.
However, numerical investigations show convincingly (but not rigorously)
that in parallel with \eqref{lowid},
 \be
\sum_{k=1}^6 g(6n+k) = 6\rho^2.
 \ee
  This in turn implies, if supplemented by a mild assumption of decay of
$g$ (after truncation), that the variance of the number of particles in a
large box will be, to leading order, the same as in the Bernoulli
measure: $v=\rho(1-\rho)$.

\section{Further remarks}\label{remarks}

1. The proofs of many of the results stated here make use of a height or
interface representation of the configurations \cite{gls}: for a
configuration $\eta$ the height $h(j)$ at site $j$ is specified, up to an
overall additive constant, by
 \be\label{eta-h}
h(j)-h(j-1)=(-1)^{\eta(j)}.
 \ee
 One can then define a dynamics for $h$ which yields the F-TASEP dynamics
for the corresponding particle configuration and is such that $h$
increases (or stays the same) with time.  When this dynamics is modified
in the high and intermediate density regions, in parallel with the
particle dynamics, $h(j)$ has a $t\to\infty$ limit in all density
regions.

 \smallskip\noindent
 2. We can consider the continuous time F-TASEP in which particles jump,
independently and at random times, to their right, provided they have an
occupied site to their left and an empty site to their right. 
The TIS state for $\rho\le1/2$ (started from $\murho$) is the same
as for the F-TASEP.  We can also consider the continuous time facilitated
{\it symmetric} exclusion, or CLG, model.  Here too the lower density
TIS state ($\rho<1/2$) is a trapped one \cite{gls2}.  The same
argument as before shows that the double zeros in the TIS state
form a renewal process.  Unlike the F-TASEP, however, this system has no
transition at $\rho=2/3$, or indeed at any value $\rho>1/2$.

 \smallskip\noindent
 3. Define the ${\rm F}_k$-TASEP by requiring that a particle have $k$
adjacent particles to its left before it can jump.  Then from the
analogue of \eqref{shel},
 \be\label{kshel}
0\,\rightarrow\,1\,\rightarrow\,\cdots\,\rightarrow\,
   k\,\leftrightarrows\,k+1\,\leftarrow\,k+2\,\leftarrow
   \,k+3\,\leftarrow\,\cdots,
 \ee
 there will again be three classes of TIS states: (a)~the low density
phase $0<\rho<k/(k+1)$, (b)~the intermediate density phase
$k/(k+1)<\rho<(k+1)/(k+2)$, and (c)~the high density phase
$(k+1)/(k+2)<\rho<1$.  The corresponding $k$-TASM will have $\n(j)\le k$
in the low density region, $\n(j)= k$ or $k+1$ in the intermediate
density region, and $\n(j)\ge k+1$ in the high density region.  We expect
that if started from a Bernoulli measure at density $\rho$ the system
will approach a stationary state.

 \par\smallskip\noindent
 4. One can show \cite{gls} that for any initial configuration $\eta_0$
with density $\rho<1/2$ there is a final configuration $\eta_\infty$ such
that $\eta_t\to\eta_\infty$ as $t\to\infty$, and that the map
$\eta_0\mapsto\eta_\infty$ commutes with translations.  The existence of
such a deterministic mapping from initial to final configurations
immediately implies that if the system is started in a TI initial state
which is ergodic, or even mixing, then the final state will be ergodic or
mixing, respectively.  Moreover, if the initial state is the Bernoulli
measure $\murho$ then the final state will in fact be isomorphic to a
Bernoulli shift, though one of lower entropy than $\murho$
\cite{ornstein}.  The same conclusions hold in the intermediate or high
density regions (but not at $\rho=1/2$ or $rho=2/3$) if one adopts the
corresponding modified dynamics discussed above.

 \par\smallskip\noindent
 5. Our analysis in this note was confined mostly to translation invariant
states.  Suppose however that we start with an initial state which
differs from the Bernoulli product state in some fixed region $[-M,M]$;
e.g., we might use there a product measure with a different density.
Then, contrary to what might be expected, at low density the resulting
stationary state will remain non-translation invariant.  At
intermediate and higher densities the limiting states will be TI, with
the localized perturbation transported away to infinity.

 \medskip\noindent
 {\bf Acknowledgments:} The work of JLL was supported by the AFOSR under award 
number FA9500-16-1-0037.

\end{document}